\documentclass[letterpaper, 10 pt, conference]{ieeeconf}

\IEEEoverridecommandlockouts
\usepackage{amsmath}
\usepackage{amssymb}
\usepackage{graphicx}
\usepackage{cite}
\usepackage{placeins}

\usepackage{amsthm}
\usepackage{xcolor}

\newcommand{\rev}[1]{#1}
\makeatletter
\def\bstctlcite{\@ifnextchar[{\@bstctlcite}{\@bstctlcite[@auxout]}}
\def\@bstctlcite[#1]#2{\@bsphack
  \@for\@citeb:=#2\do{%
    \edef\@citeb{\expandafter\@firstofone\@citeb}%
    \if@filesw\immediate\write\csname #1\endcsname{\string\citation{\@citeb}}\fi}%
  \@esphack}
\makeatother
\usepackage{soul}
\sethlcolor{yellow}

\theoremstyle{plain}

\theoremstyle{definition}
\newtheorem{definition}{Definition}
\theoremstyle{remark}

\newif\ifdraftnotes
\draftnotesfalse

\title{\LARGE \bf
Closed-Loop Refinement and Execution for Learned Driving Planners
}

\author{
Huaijin Hu$^{1}$,
Shanting Wang$^{1}$,
Zhongyu Mo$^{1}$,
and Andreas A. Malikopoulos$^{2}$,
\textit{Senior Member, IEEE}
\thanks{This research was supported in part by NSF under Grants
CNS-2401007, CMMI-2348381, IIS-2415478, and in part by MathWorks.}%
\thanks{$^{1}$Huaijin Hu, Shanting Wang, and Zhongyu Mo are with the
Systems Engineering Program, Cornell University, Ithaca, NY 14853 USA
(emails: \texttt{hh833@cornell.edu};
\allowbreak\texttt{sw997@cornell.edu};
\allowbreak\texttt{zm73@cornell.edu}).}
\thanks{$^{2}$Andreas A. Malikopoulos is with the Applied Mathematics, Systems Engineering, Robotics, Mechanical Engineering, Electrical and Computer Engineering, and Civil \& Environmental Engineering, Cornell University, Ithaca, NY, USA. (email: \texttt{amaliko@cornell.edu}).}%
}

\begin{document}
\bstctlcite{IEEEexample:BSTcontrol}

\maketitle
\thispagestyle{empty}
\pagestyle{empty}

\begin{abstract}
Learning-based driving planners are usually trained and evaluated in open loop
against logged trajectories. In closed loop, a trajectory with small
displacement error can still stall the vehicle, steer it into a conflict
with surrounding agents, or be executed with abrupt braking. We introduce Closed-Loop
Refinement and Execution (CLRE), a hierarchical receding-horizon control
framework designed to mitigate these failure modes while leaving the upstream
planner frozen and adding no new learned model. The upper layer treats the
nominal trajectory as a reference and solves a finite-horizon optimal control
problem that trades route progress against interaction with predicted agents.
Solving it from several initializations gives a candidate set, and a prediction-conditioned oriented-bounding-box (OBB) feasibility test retains
only candidates whose minimum predicted OBB clearance over the horizon meets a
threshold. The lower layer
executes the lowest-cost survivor, or a route-centerline backup when none
remains, through the tracking controller supplied with the planner, augmented
by a range-based speed bound and a saturated proportional braking law. In closed-loop simulation on the 220-route Bench2Drive validation set with VAD as the upstream planner, CLRE raises the driving score from 42.26 to 55.89
and route completion from 55.69 to 71.51, and reduces collision events from
117 to 97.
\end{abstract}

\section{Introduction}
\label{sec:intro}
End-to-end autonomous driving has emerged as an alternative to
conventional modular pipelines by learning driving decisions directly
from sensor observations and scene representations. Methods such as
UniAD~\cite{hu2023planning} and VAD~\cite{jiang2023vad} predict a
short-horizon ego trajectory that is subsequently converted into
vehicle commands by a downstream tracking controller. These planners
are commonly trained and evaluated in open loop against logged expert
trajectories. However, accurate open-loop trajectory prediction does
not necessarily imply reliable closed-loop execution.

In closed loop each executed action changes the next observation, so
planning errors accumulate. Open-loop and closed-loop performance can be
substantially misaligned~\cite{dauner2023parting}, Li \emph{et
al.}~\cite{li2024egostatus} showed that open-loop displacement metrics can
be largely explained by the ego status alone, and
NeuroNCAP~\cite{ljungbergh2024neuroncap} reported that planners with strong
open-loop performance still collide in interactive scenarios. Bench2Drive
makes the point sharper by testing end-to-end systems in closed-loop CARLA
scenarios, where collisions, stalling, and route incompletion all count
against the score~\cite{jia2024bench2drive}. In this work, we focus on three
practically important failure modes that arise when learned trajectories are
executed in closed loop: insufficient route progress or stalling, motion
toward predicted conflicts, and abrupt longitudinal intervention.

Several lines of work try to improve closed-loop planning.
One line modifies the learned planner itself, for example
through a kinematic refinement network~\cite{li2025hydranext} or online
test-time adaptation~\cite{sima2025centaur}. Such approaches improve
closed-loop behavior but modify or adapt the learned planning stack.
Another line explicitly optimizes or refines trajectories at deployment.
Examples include coupling learned prediction with differentiable
planning~\cite{huang2024dipp}, tracking learned waypoints with model predictive
control (MPC)~\cite{zhou2024fusionassurance}, handling multimodal prediction
uncertainty through chance constraints~\cite{nair2023smpc}, test-time
trajectory search with a frozen learned scorer~\cite{xu2026toad}, and
finite-horizon optimization with interaction-aware geometric
representations~\cite{wang2025visiopath,wang2025corra}.
These methods improve trajectories through optimization, with interaction
and safety handled through objective terms, probabilistic constraints, or
method-specific verification mechanisms.
Runtime monitors can instead provide a separate check on a learned plan and
invoke fallback when necessary~\cite{vitelli2022safetynet}. Such monitoring
can reject an unsafe nominal plan, but does not by itself generate alternative
refined candidates. 

Supervising a nominal policy with a separate safety layer is well established
in control. Predictive safety filters, control barrier functions, and
reachability-based methods modify or override nominal inputs to maintain
prescribed safety
conditions~\cite{hewing2020learning,wabersich2023safetyfilters}, and barrier
certificates~\cite{ChalakiCBF2022} and safety-prioritized receding-horizon
control~\cite{mahbub2023_automatica} have been applied to vehicle coordination.
These methods check safety against an explicit model of the environment. A learned driving
stack instead supplies its own predictions of surrounding agents, so any
runtime condition built on them inherits their limitations.

\rev{These gaps motivate our approach. We propose the Closed-Loop Refinement and
Execution (CLRE) framework, which keeps the upstream planner frozen and adds no
new learned model. Unlike waypoint-tracking MPC or runtime filters that operate
on a single nominal plan, CLRE first refines the plan to recover progress and
reduce predicted interactions, and then independently screens the resulting
candidates with a prediction-conditioned oriented-bounding-box(OBB) feasibility test. The lower layer
executes the selected candidate with additional intervention when needed.}

The main contributions of this paper are the following.
\begin{itemize}
\item We propose CLRE, a training-free framework that improves the
closed-loop behavior of a frozen learned planner through trajectory
refinement, OBB feasibility testing, and execution-level intervention.
\item We develop a receding-horizon refinement method that balances
nominal-path tracking, route progress, and predicted-agent interaction,
followed by an OBB feasibility test that screens the resulting candidates
against predicted agent motion.
\item We introduce an execution layer with a forward-range speed bound,
backup policy, and saturated braking, and evaluate the complete framework
on Bench2Drive through ablation, sensitivity, and cross-planner studies.
\end{itemize}

The remainder of this paper is organized as follows:
Section~\ref{sec:problem} formulates the receding-horizon refinement problem
and Section~\ref{sec:method} constructs the objective, the OBB feasibility test,
and the execution layer. Section~\ref{sec:experiments} reports closed-loop
benchmark results. Finally, we draw concluding remarks in
Section~\ref{sec:conclusion}.
\section{Problem Formulation}
\label{sec:problem}
\subsection{Vehicle Model and Constraints}
\label{sec:problem:system}

At each closed-loop replanning instant, indexed by $t\in\mathbb{N}$, CLRE
optimizes the ego motion over a horizon of $N$ intervals with time step
$\Delta t$. The prediction index $k\in\{0,\ldots,N\}$ denotes the number of
steps ahead of the current replanning instant, so that $k\Delta t$ is the
corresponding prediction time relative to $t$.

\begin{figure}[!tb]
    \centering
    \includegraphics[width=0.7\columnwidth]{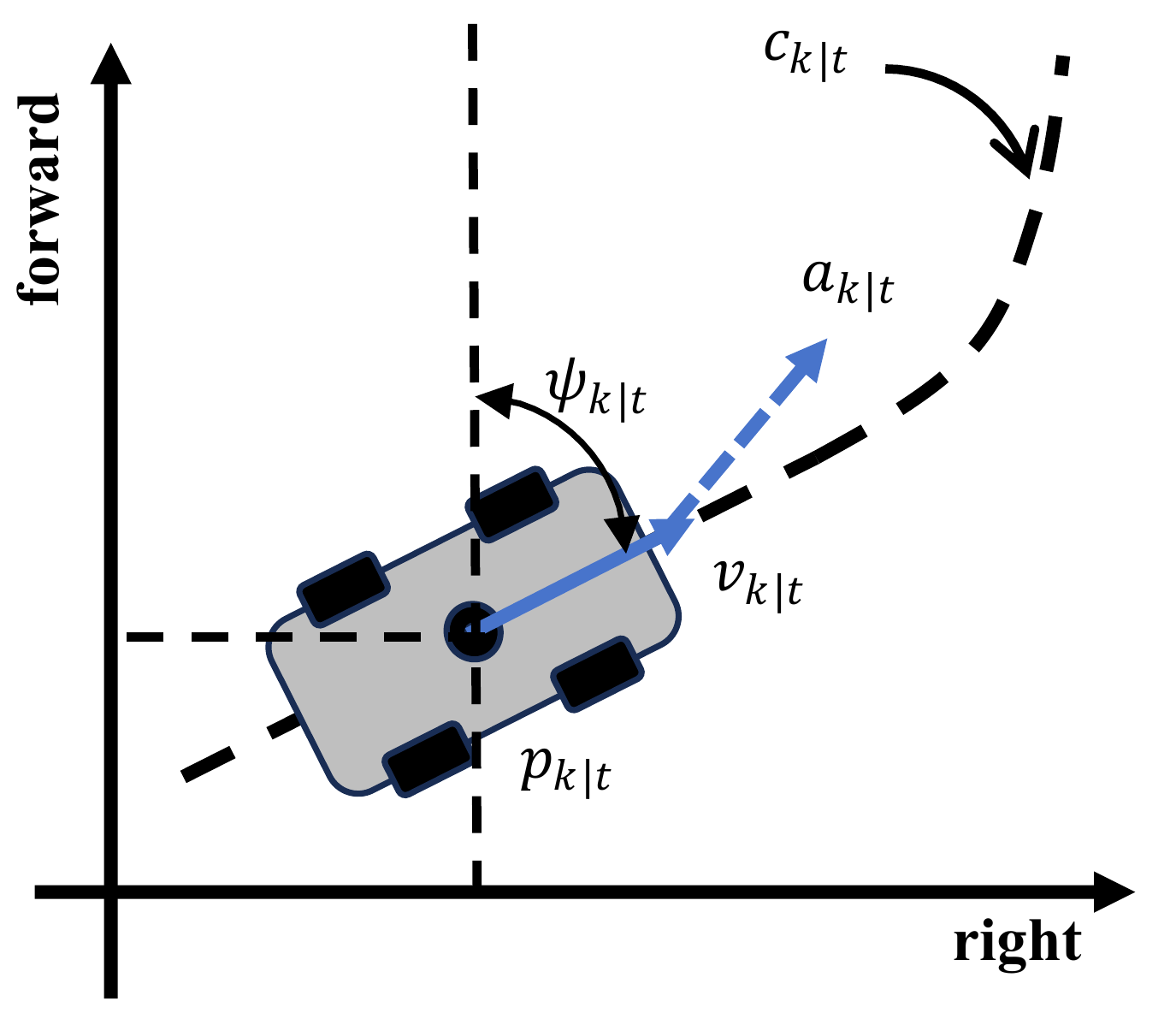}
    \caption{Local coordinate frame, ego state, and control variables used
    for trajectory refinement. The frame is fixed at the ego pose at
    replanning instant $t$. The positive position axes point rightward and
    forward, and positive curvature corresponds to a right turn.}
    \label{fig:state_dynamics}
\end{figure}

All planar positions, headings, and geometric scene information used in the
refinement problem are represented in a local coordinate frame
$\mathcal{F}_t$ fixed at the current ego pose. As shown in
Fig.~\ref{fig:state_dynamics}, the first position coordinate points to the
right of the vehicle and the second points forward. The heading angle is
measured clockwise from the forward axis.

The predicted ego state is
\[
x_{k|t}
=
\begin{bmatrix}
p_{k|t}^{\top} & \psi_{k|t} & v_{k|t}
\end{bmatrix}^{\top}
\in\mathbb{R}^{4},
\]
where $p_{k|t}\in\mathbb{R}^{2}$ is the planar ego position in $\mathcal{F}_t$,
$\psi_{k|t}$ is the heading angle, and $v_{k|t}$ is the longitudinal speed. Since $\mathcal{F}_t$ is fixed at the
current ego pose, the initial state satisfies
\begin{equation}
p_{0|t}=[0\;\,0]^{\top},
\qquad
\psi_{0|t}=0,
\qquad
v_{0|t}=v_t,
\label{eq:initial_state}
\end{equation}
where $v_t$ is the measured ego speed at replanning instant $t$.

The control input is $u_{k|t}=[a_{k|t}\;\,c_{k|t}]^{\top}\in\mathbb{R}^{2}$,
where $a_{k|t}$ is the longitudinal acceleration and $c_{k|t}$ is the path
curvature. For a kinematic bicycle model with wheelbase $L$ and steering
angle $\delta$, the path curvature satisfies
$c=\tan(\delta)/L$~\cite{polack2017kinematic}, with the steering angle
defined positive to the right. Instead of deriving the curvature bound from a
steering limit, we impose a prescribed curvature limit $c_{\max}$ directly,
which corresponds to a minimum turning radius of $1/c_{\max}$.

The admissible state and input constraints are
\begin{equation}
\begin{aligned}
0 &\leq v_{k|t}\leq v_{\max},
&& k=0,\ldots,N,\\
a_{\min} &\leq a_{k|t}\leq a_{\max},
&& k=0,\ldots,N-1,\\
|c_{k|t}| &\leq c_{\max},
&& k=0,\ldots,N-1,
\end{aligned}
\label{eq:system_constraints}
\end{equation}
where $v_{\max}$ is the maximum longitudinal speed and
$a_{\min}$ and $a_{\max}$ are the minimum and maximum longitudinal
accelerations.

Using curvature as the lateral input, we describe the ego motion by the
continuous-time kinematic model $\dot v=a$, $\dot\psi=vc$, and
$\dot p=v[\sin\psi\;\,\cos\psi]^{\top}$, which is discretized over the
refinement interval $\Delta t$ using a semi-implicit Euler scheme:
\begin{equation}
\begin{aligned}
v_{k+1|t} &= v_{k|t}+a_{k|t}\Delta t,\\
\psi_{k+1|t} &= \psi_{k|t}+v_{k+1|t}c_{k|t}\Delta t,\\
p_{k+1|t} &= p_{k|t}+v_{k+1|t}\Delta t
\begin{bmatrix}
\sin\psi_{k+1|t}\\
\cos\psi_{k+1|t}
\end{bmatrix},
\end{aligned}
\label{eq:system_dynamics}
\end{equation}
for $k=0,\ldots,N-1$.

\begin{definition}[Nominal trajectory]
\label{def:nominal}
At replanning instant $t$, the frozen learned planner produces the nominal
trajectory
$\tau_t^{\mathrm{nom}}=(p_{1|t}^{\mathrm{nom}},\ldots,p_{M|t}^{\mathrm{nom}})$,
where $p_{j|t}^{\mathrm{nom}}\in\mathbb{R}^{2}$ is the planner-predicted ego position at
relative prediction time $j\Delta t_{\mathrm{nom}}$, expressed in
$\mathcal{F}_t$, and $\Delta t_{\mathrm{nom}}$ is the planner waypoint
interval. The nominal trajectory serves as a reference for refinement and is
not assumed to satisfy the vehicle dynamics~\eqref{eq:system_dynamics} or the
state and input constraints~\eqref{eq:system_constraints}.
\end{definition}

At each replanning instant, CLRE also receives the surrounding-agent motion
predictions provided by the frozen upstream planner, the local lane-boundary
geometry, and the navigation route. We denote this information collectively
by $\mathcal{E}_t$. All information in $\mathcal{E}_t$ is available at
replanning instant $t$ and future ground-truth agent states are not used.
\subsection{Receding-Horizon Refinement Problem}
\label{sec:problem:optimization}

The nominal and refined trajectories span the same prediction horizon. The
refinement interval $\Delta t$ is chosen such that
$\Delta t_{\mathrm{nom}}=q\Delta t$ with $q\in\mathbb{N}_{>0}$, and therefore
$N=qM$ and $N\Delta t=M\Delta t_{\mathrm{nom}}$.

Define the predicted state and control sequences as
$\mathbf{x}_t=(x_{0|t},\ldots,x_{N|t})$ and
$\mathbf{u}_t=(u_{0|t},\ldots,u_{N-1|t})$. For a given control sequence $\mathbf{u}_t$, the corresponding state sequence
$\mathbf{x}_t$ is generated recursively from
\eqref{eq:initial_state} and \eqref{eq:system_dynamics}.

The refinement objective depends on the predicted ego trajectory, the control
sequence, the nominal planner output, and the environment information
available at the replanning instant $t$. For compactness, we write
\begin{equation}
J_t(\mathbf{x}_t,\mathbf{u}_t)
:=
J\!\left(
\mathbf{x}_t,\mathbf{u}_t;
\tau_t^{\mathrm{nom}},\mathcal{E}_t
\right),
\label{eq:objective_dependency}
\end{equation}
with its individual terms defined in Section~\ref{sec:method}.

The finite-horizon trajectory-refinement problem is
\begin{equation}
\begin{aligned}
\min_{\mathbf{u}_t}\quad
& J_t(\mathbf{x}_t,\mathbf{u}_t)
\\
\text{s.t.}\quad
& \eqref{eq:initial_state},\
  \eqref{eq:system_constraints},\
  \eqref{eq:system_dynamics}.
\end{aligned}
\label{eq:refinement_ocp}
\end{equation}

For a candidate control sequence $\mathbf{u}_t$ and its resulting state
trajectory $\mathbf{x}_t$, the corresponding refined position trajectory is
$\tau_t^{\mathrm{ref}}=(p_{1|t},\ldots,p_{N|t})$. After candidate selection, the selected refined trajectory is downsampled to
the waypoint rate expected by the downstream tracking controller. Since
$\Delta t_{\mathrm{nom}}=q\Delta t$, the resulting reference sequence is
$\tau_t^{\mathrm{exec}}=(p_{q|t},p_{2q|t},\ldots,p_{Mq|t})$. This waypoint sequence is passed to the execution layer of
Section~\ref{sec:method:controller}.
\section{Solution Approach}
\label{sec:method}
\subsection{Overview}
\label{sec:method:overview}

\begin{figure*}[t]
    \vspace*{6pt}
    \centering
    \includegraphics[width=\textwidth]{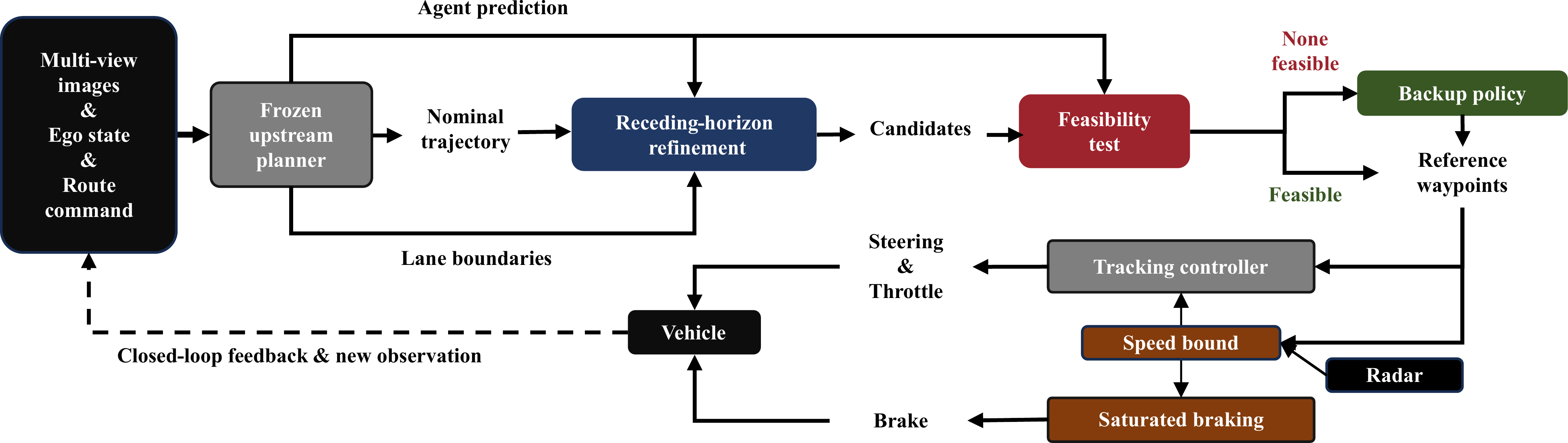}
    \caption{Overview of CLRE. Gray blocks are supplied unchanged with the
    upstream planner, colored blocks are introduced by CLRE, and dark blocks
    denote sensor inputs and the vehicle.}
    \label{fig:arch}
\end{figure*}

The overall CLRE architecture is shown in Fig.~\ref{fig:arch}. The following
subsections construct the refinement objective, candidate-selection test, and
execution layer used to solve and execute the refinement problem of
Section~\ref{sec:problem:optimization}.

\subsection{Trajectory-Level Terms}
\label{sec:method:trajectory}

Let $r_t\in\mathbb{R}^2$ denote the route target point provided by the upstream route planner,
expressed in $\mathcal{F}_t$. The local route direction is
\begin{equation}
g_t
=
\begin{cases}
\dfrac{r_t}{\|r_t\|_2},
& \|r_t\|_2>1~\mathrm{m},\\[8pt]
\begin{bmatrix}0 & 1\end{bmatrix}^{\top},
& \text{otherwise},
\end{cases}
\qquad
\|g_t\|_2=1,
\label{eq:route_direction}
\end{equation}
and its unit normal is
$n_t=[g_{t,2},-g_{t,1}]^\top$.

\paragraph{Point-wise Nominal Tracking}

Let $\tilde p_{k|t}^{\mathrm{nom}}$ denote the piecewise-linear interpolation
of
$p_{0|t},p_{1|t}^{\mathrm{nom}},\ldots,p_{M|t}^{\mathrm{nom}}$
at relative time $k\Delta t$. We define
\begin{equation}
j_{\mathrm{pt}}
=
\sum_{k=1}^{N}
H_{\delta_{\mathrm H}}
\left(
\left\|
p_{k|t}-\tilde p_{k|t}^{\mathrm{nom}}
\right\|_2
\right),
\label{eq:pt}
\end{equation}
where $H_{\delta_{\mathrm H}}(\cdot)$ is the Huber loss with transition
threshold $\delta_{\mathrm H}$. This term anchors the refinement to the
time-parameterized motion proposed by the learned planner, thereby retaining
its local driving intent. The Huber loss penalizes small deviations
quadratically but grows only linearly for larger deviations, allowing the
refinement to depart from the nominal trajectory when the other objective
terms favor a substantially different motion.

\paragraph{Geometric Path Commitment}

Point-wise tracking constrains both where the ego travels and when it reaches
each nominal waypoint. To preserve the nominal path geometry without imposing
the same timing, we additionally define
\begin{equation}
j_{\mathrm{ct}}
=
\gamma_t
\sum_{k=1}^{N}
\min_{\zeta\in\mathcal P_t^{\mathrm{nom}}}
\left\|p_{k|t}-\zeta\right\|_2 ,
\label{eq:ct}
\end{equation}
where $\mathcal P_t^{\mathrm{nom}}$ is the nominal path sampled at $0.4$~m
arc-length intervals and extended by $45$~m, which equals $Nv_{\max}\Delta t$,
along its terminal tangent. If the nominal path is too short to define a
terminal tangent, $g_t$ is used instead. The extension prevents the finite
nominal endpoint from acting as an artificial stopping point.

\rev{The commitment gate is
$\gamma_t = [1+\exp(-\beta_\gamma(s_t-\bar s))]^{-1}$, where
$s_t=(p_{M|t}^{\mathrm{nom}}-p_{0|t})^\top g_t$ is the nominal forward
displacement along the route direction. $\beta_\gamma>0$ controls the
steepness of the gate, and $\bar s$ is the displacement at which
$\gamma_t=1/2$.} When the nominal
trajectory exhibits little forward progress, $\gamma_t$ decreases, reducing
the influence of a short or stalled nominal path on the refinement.

\paragraph{Forward Progress}

The preceding terms retain information from the nominal planner but do not
themselves encourage the vehicle to advance when that planner becomes overly
conservative or stalls. We therefore introduce
\begin{equation}
j_{\mathrm{prog}}
=
-\left(
p_{N|t}-p_{0|t}
\right)^{\top}g_t .
\label{eq:prog}
\end{equation}
Minimizing $j_{\mathrm{prog}}$ rewards terminal displacement in the route
direction. Since $\|g_t\|_2=1$ and $v_{k|t}\leq v_{\max}$ under \eqref{eq:system_constraints},
$|j_{\mathrm{prog}}|\leq Nv_{\max}\Delta t$.

\subsection{Interaction-Level Terms}
\label{sec:method:interaction}

Let $\mathcal I_t$ denote the set of surrounding agents retained from the
upstream planner. For each $i\in\mathcal I_t$, CLRE uses the
highest-probability motion mode, linearly interpolated onto the refinement
grid from the agent's current position. \rev{An offline check on the development windows showed that enforcing
clearance against all modes with probability at least $0.1$ increased the
fraction of windows with no feasible candidate from $11.2\%$ to $22.6\%$.
We therefore retain top-1 screening in the current implementation, while the
confidence $\rho_i$ modulates the soft interaction cost in~\eqref{eq:interaction}.
Uncertainty-aware multimodal screening is left to future work.}

\begin{figure}[!tb]
    \centering
    \includegraphics[width=0.95\columnwidth]{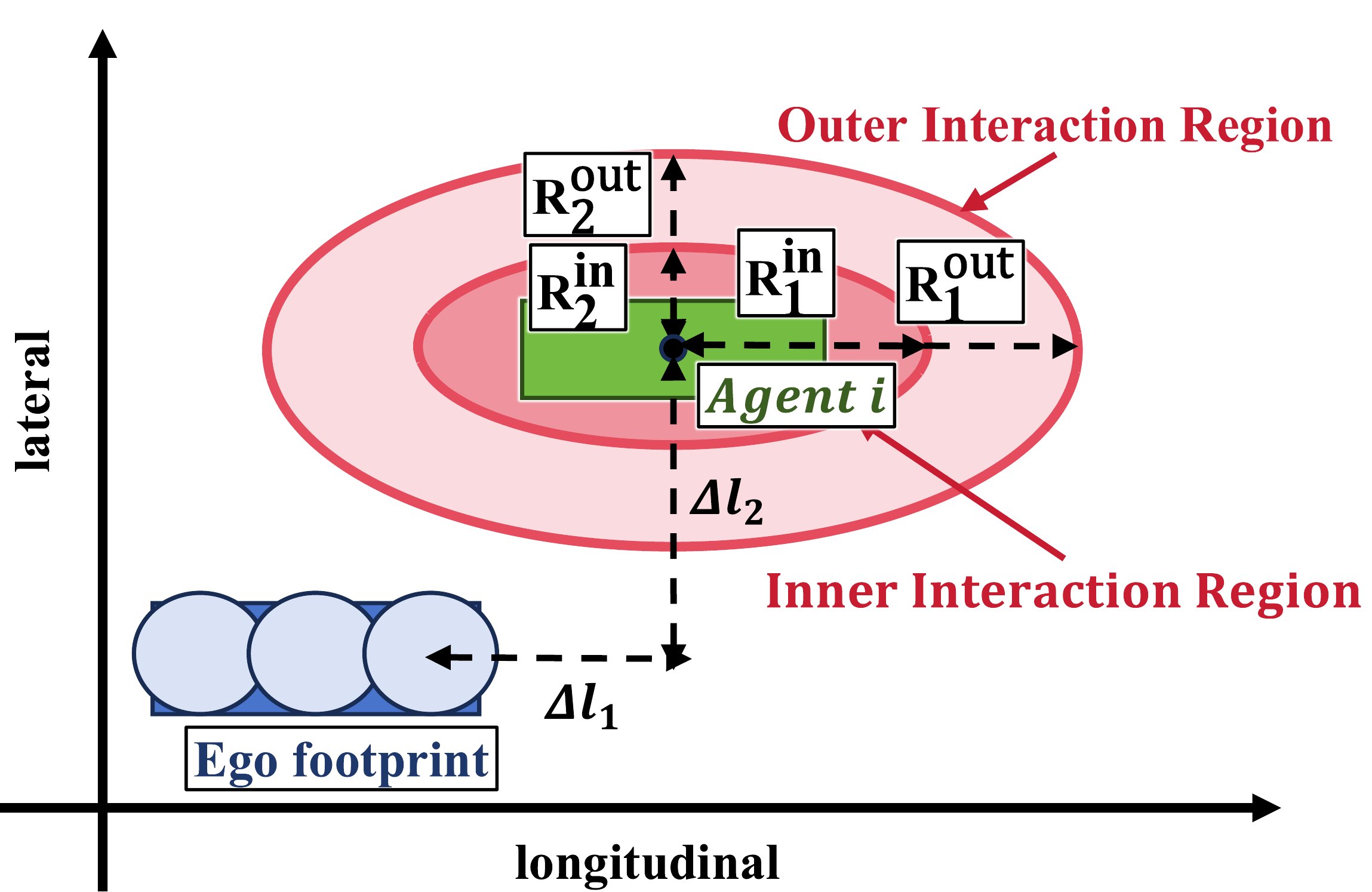}
    \caption{Outer and inner interaction regions around a predicted agent.}
    \label{fig:ellipse_interaction}
\end{figure}

During optimization, the ego footprint is represented by three overlapping
disks distributed along its longitudinal axis. Compared with a single
centered disk, this representation captures the longitudinal extent of the
vehicle more accurately while retaining a simple distance-based geometry for
optimization. For each predicted agent, outer and inner ellipses define the
interaction regions shown in Fig.~\ref{fig:ellipse_interaction}.

For ego disk $d$, let $\Delta l_{1,k|t}^{i,d}$ and
$\Delta l_{2,k|t}^{i,d}$ denote the longitudinal and lateral relative
coordinates in the agent-aligned frame. For
$\sigma\in\{\mathrm{out},\mathrm{in}\}$, define
\begin{equation}
\eta_{k|t}^{i,\sigma}
=
\min_{d\in\{1,2,3\}}
\sqrt{
\left(
\frac{\Delta l_{1,k|t}^{i,d}}
     {R_{1}^{i,\sigma}}
\right)^2
+
\left(
\frac{\Delta l_{2,k|t}^{i,d}}
     {R_{2}^{i,\sigma}}
\right)^2
},
\label{eq:ellipse_distance}
\end{equation}
where $R_{1}^{i,\sigma}$ and $R_{2}^{i,\sigma}$ are the longitudinal and
lateral ellipse radii. Hence, $\eta_{k|t}^{i,\sigma}=1$ corresponds to the
ellipse boundary, $\eta_{k|t}^{i,\sigma}>1$ to separation, and
$\eta_{k|t}^{i,\sigma}<1$ to penetration of the interaction region by at
least one ego disk.

We use $\phi_\beta(z)=\beta^{-1}\log(1+\exp(\beta z))$
as a smooth approximation of $[z]_+=\max(z,0)$. In the interaction terms below, $z=1-\eta$. The penalty is
therefore small outside an interaction ellipse and increases continuously as
the ego footprint moves inside it.

\paragraph{Predicted-Agent Interaction}

Let $\rho_i$ be the probability assigned to the selected motion mode of
agent $i$. The interaction cost is
\begin{equation}
\begin{aligned}
j_{\mathrm{int}}
=
\sum_{i\in\mathcal I_t}
\rho_i
\sum_{k=1}^{N}
\Big[
&
w_{\mathrm{out}}
\phi_{\beta_{\mathrm{out}}}
\left(
1-\eta_{k|t}^{i,\mathrm{out}}
\right)
\\
&+
w_{\mathrm{in}}
\phi_{\beta_{\mathrm{in}}}
\left(
1-\eta_{k|t}^{i,\mathrm{in}}
\right)
\Big].
\end{aligned}
\label{eq:interaction}
\end{equation}
The weights $w_{\mathrm{out}},w_{\mathrm{in}}\geq0$ set the relative
importance of the two interaction regions, while
$\beta_{\mathrm{out}},\beta_{\mathrm{in}}>0$ control the sharpness of their
softplus penalties. The outer ellipse introduces an anticipatory penalty before the ego approaches
the agent closely, whereas the tighter inner ellipse penalizes near-contact
motion more strongly. Together, they shape candidate trajectories away from
predicted conflicts without imposing a hard collision constraint inside the
refinement problem. The confidence weight $\rho_i$ is not used in the feasibility test of Section~\ref{sec:method:safety}.

\paragraph{Road-Boundary Proximity}

Let $\{b_\ell\}_{\ell=1}^{N_b}$ denote sampled points on the retained solid
lane boundaries. We define
\begin{equation}
j_{\mathrm{bnd}}
=
\sum_{k=1}^{N}
\phi_{\beta_{\mathrm{bnd}}}
\left(
m_{\mathrm{bnd}}
-
\min_{\ell=1,\ldots,N_b}
\left\|
p_{k|t}-b_\ell
\right\|_2
\right),
\label{eq:bnd}
\end{equation}
where $m_{\mathrm{bnd}}>0$ is the prescribed boundary-proximity margin and
$\beta_{\mathrm{bnd}}>0$ controls the sharpness of the penalty. The cost
increases as the refined trajectory enters the prescribed margin,
discouraging motion toward solid boundaries and non-drivable regions.

\paragraph{Lateral Corridor Regulation}

The road-boundary term depends on the locally available boundary geometry.
We additionally use a route-aligned corridor to limit large lateral
departures from the navigation direction. The lateral displacement is
$e_{k|t}=|(p_{k|t}-p_{0|t})^{\top}n_t|$, with corresponding cost
\begin{equation}
j_{\mathrm{cor}}
=
\sum_{k=1}^{N}
\max
\left(
e_{k|t}-\bar y,\,
0
\right)^2,
\label{eq:cor}
\end{equation}
where $\bar y$ is the corridor half-width and deviations beyond it are increasingly penalized.
This term complements the explicit boundary cost by preventing the optimizer
from exploiting large lateral excursions in pursuit of route progress.

\subsection{Ego-Motion-Level Terms}
\label{sec:method:ego}

\paragraph{Speed Regularization}

Let $v_t^{\mathrm{nom}}$ denote the speed reference obtained from the mean
spacing of consecutive nominal waypoints divided by
$\Delta t_{\mathrm{nom}}$, restricted to a prescribed speed interval. We
define
\begin{equation}
j_v
=
\sum_{k=1}^{N}
\max
\left(
v_{k|t}-v_t^{\mathrm{nom}},
0
\right)^2 .
\label{eq:speed}
\end{equation}
This one-sided penalty prevents the optimizer from obtaining additional route
progress simply by driving substantially faster than the nominal plan.
Speeds below $v_t^{\mathrm{nom}}$ are not penalized, allowing the refinement
to slow the vehicle when required by predicted interactions or road geometry.

\paragraph{Input Regularization}

To discourage rapidly varying control profiles and unnecessarily sharp
steering, we define
\begin{equation}
j_{\mathrm{sm}}
=
\sum_{k=1}^{N-1}
\left[
\left(\Delta a_{k|t}\right)^2
+
\lambda_c
\left(\Delta c_{k|t}\right)^2
\right]
+
\lambda_0
\sum_{k=0}^{N-1}
c_{k|t}^{2},
\label{eq:smoothness}
\end{equation}
where
$\Delta a_{k|t}=a_{k|t}-a_{k-1|t}$ and
$\Delta c_{k|t}=c_{k|t}-c_{k-1|t}$,
and $\lambda_c,\lambda_0\geq0$ weight curvature variation and curvature
magnitude, respectively.

\subsection{Solution of the Refinement Problem}
\label{sec:method:optimization}

Combining the preceding terms, the refinement objective introduced in
\eqref{eq:objective_dependency} is
\begin{equation}
\begin{aligned}
J_t ={}&
w_{\mathrm{pt}}j_{\mathrm{pt}}
+
w_{\mathrm{ct}}j_{\mathrm{ct}}
+
w_{\mathrm{prog}}j_{\mathrm{prog}}
+
j_{\mathrm{int}}
\\
&+
w_{\mathrm{bnd}}j_{\mathrm{bnd}}
+
w_{\mathrm{cor}}j_{\mathrm{cor}}
+
w_vj_v
+
w_{\mathrm{sm}}j_{\mathrm{sm}} .
\end{aligned}
\label{eq:cost_function}
\end{equation}
The weights are nonnegative, fixed offline, and reported in
Section~\ref{sec:sensitivity}.

Problem~\eqref{eq:refinement_ocp} is nonconvex and piecewise
differentiable, and we generate candidate trajectories using
Adam~\cite{kingma2015adam}. In the numerical implementation, the
acceleration and curvature variables are clamped to their admissible bounds
before each rollout, and the speed is projected onto $[0,v_{\max}]$ after
each Euler update. These projections serve as numerical safeguards for
constraint handling.

Eight behavior-structured initializations are
used to explore different local solutions, including nominal-following,
braking, progress-oriented, accelerate-then-coast, and lateral-avoidance
behaviors. When available, the control sequence selected at the
previous replanning instant is used as an additional warm-start
initialization, shifted forward by one refinement step each time an
interval $\Delta t$ has elapsed. Its unoptimized shifted
version is also retained for candidate evaluation.

Because Adam is run for a fixed number of iterations on a nonconvex
objective, we claim that each solve produces a locally refined candidate but not the exact global solution. The numerical
sensitivity of candidate selection and feasibility to larger iteration
budgets is evaluated in Section~\ref{sec:solver_analysis}.

\subsection{Prediction-Conditioned OBB Feasibility Test}
\label{sec:method:safety}

Because agent interaction enters the refinement objective through soft
penalties, each candidate is additionally screened using oriented bounding
boxes (OBBs). For a given candidate, let $\mathcal B_{k|t}^{\mathrm{ego}}$
and $\mathcal B_{k|t}^{i}$ denote the ego OBB and the predicted OBB of agent
$i$ at step $k$, respectively.

We define
$d_{\mathrm{OBB}}(\mathcal B_1,\mathcal B_2)$ as the signed box-to-box
clearance: positive for separated boxes and negative for overlapping boxes.
For separated boxes, the Euclidean box-to-box clearance is computed as the
minimum distance from the corners of each box to the other box, which is exact
for rectangles. For overlapping boxes, the penetration depth is computed using
the separating-axis construction~\cite{ericson2005realtime}.

The minimum predicted clearance of a candidate over the refinement horizon
is
\begin{equation}
\operatorname{gap}
=
\min_{\substack{
i\in\mathcal I_t\\
k=1,\ldots,N
}}
d_{\mathrm{OBB}}
\left(
\mathcal B_{k|t}^{\mathrm{ego}},
\mathcal B_{k|t}^{i}
\right).
\label{eq:obb_gate}
\end{equation}
A candidate is feasible if $\operatorname{gap}\geq\epsilon$, where $\epsilon>0$ is the prescribed clearance threshold. Among
feasible candidates, the one with the smallest $J_t$ is selected. If a completed refinement solve returns no feasible candidate, the backup
policy of Section~\ref{sec:method:controller} is used. \rev{As the test is evaluated against predicted agent motion, feasibility is prediction-conditioned rather than a formal safety certificate, which means prediction errors can invalidate the clearance estimate. CLRE reduces this exposure through receding-horizon replanning and the prediction-independent radar intervention in Section~\ref{sec:method:controller}.}

\subsection{Execution Layer}
\label{sec:method:controller}

The execution layer converts the selected reference waypoints into steering,
throttle, and brake commands. The baseline steering controller tracks the
reference waypoints, so the optimized acceleration and curvature sequence is
used only during refinement and is not applied directly to the vehicle. The
baseline throttle PID is retained, and longitudinal execution is augmented as
described below.

\paragraph{Backup Policy}

If a completed refinement solve returns no feasible candidate, the reference
waypoints are generated from the navigation-route centerline. The measured ego speed at entry to the
backup episode is latched and used as the longitudinal reference until a
feasible candidate becomes available.

\paragraph{Forward-Range Speed Bound}

A forward-facing radar provides a planner-independent longitudinal speed
ceiling. After filtering returns to a forward corridor, let $d_t$ denote the
range to the closest retained return and $v_t^{\mathrm{close}}$ its measured
closing rate, with
$v_t^{\mathrm{close}}>0$ indicating a closing obstacle. The speed ceiling is active only when
$v_t^{\mathrm{close}}>0$.

Using a stationary-obstacle stopping-distance approximation based on the
current measured range, we define
\[
vT_r+\frac{v^2}{2a_{\mathrm{br}}}
\leq
\max(d_t-d_{\min},0).
\]
where $a_{\mathrm{br}}>0$ is the assumed braking deceleration. The largest
admissible speed is therefore
\[
v_t^{\mathrm{safe}}
=
-a_{\mathrm{br}}T_r
+
\sqrt{
(a_{\mathrm{br}}T_r)^2
+
2a_{\mathrm{br}}\max(d_t-d_{\min},0)
}.
\]
If no retained return is closing, $v_t^{\mathrm{safe}}=+\infty$.

Let $v_t^{\mathrm{ref}}$ denote the speed implied by the selected reference
waypoints, or the latched speed during backup. The commanded speed is
$v_t^{\mathrm{cmd}}=\min(v_t^{\mathrm{ref}},v_t^{\mathrm{safe}})$.

For a closing return, the measured time to collision is
$\operatorname{TTC}_t=d_t/v_t^{\mathrm{close}}$. If it remains below a
prescribed emergency threshold for a fixed number of consecutive control
updates, full braking is applied.

\paragraph{Saturated Proportional Braking}

With
$\Delta v_t=v_t-v_t^{\mathrm{cmd}}$, the normalized brake command is
\[
u_t^{\mathrm{br}}
=
\operatorname{sat}_{[0,1]}
\left(
\frac{\Delta v_t-\Delta v_{\min}}
{\Delta v_{\max}-\Delta v_{\min}}
\right),
\]
where $\operatorname{sat}_{[0,1]}(\cdot)$ denotes saturation to $[0,1]$,
and $\Delta v_{\min}$ and $\Delta v_{\max}$ are the brake-onset and
full-brake thresholds. Full braking is also applied when
$v_t^{\mathrm{cmd}}$ falls below the stopping-speed threshold. Otherwise, the baseline longitudinal PID
controller computes the throttle command using $v_t^{\mathrm{cmd}}$ as its
target.
\section{Simulation Results}
\label{sec:experiments}
\subsection{Simulation Setup}
\label{sec:experimental_setup}

We evaluate CLRE on Bench2Drive v0.0.4~\cite{jia2024bench2drive} with CARLA
0.9.15. The main evaluation uses VAD~\cite{jiang2023vad} on all 220 official
validation routes, covering 44 different scenarios, while ablations use the first 36 routes. We report Driving Score (DS), Route Completion (RC), Success Rate (SR), and total collision events following Bench2Drive.

The refinement horizon is $3$~s, with $M=6$, $\Delta t_{\mathrm{nom}}=0.5$~s,
$N=12$, and $\Delta t=0.25$~s. We use $v_{\max}=15$~m/s,
$[a_{\min},a_{\max}]=[-6,2.5]$~m/s$^2$, $c_{\max}=0.2$~m$^{-1}$, and the OBB clearance threshold is
$\epsilon=0.05$~m. The ellipse radii are obtained by enlarging the agent half-length and
half-width by the $1.3$~m ego-footprint disk radius and by margins of
$0.6$/$0.4$~m for the outer ellipse and $0.05$~m for the inner ellipse. Adam uses a learning rate of $0.06$, with $70$ iterations
for each cold start and $25$ for the warm start.

The execution layer adds a forward radar only for the forward speed bound and the emergency braking condition. The navigation route is provided by the benchmark, and the lane
boundaries and surrounding-agent predictions are obtained from the frozen upstream planner without ground-truth future states.

Geometric and horizon parameters are determined by vehicle geometry, planner
sampling, and prescribed design margins. The objective weights and braking
thresholds are heuristic parameters chosen manually during method development,
rather than optimized against benchmark performance, and are frozen for all
reported evaluations. Section~\ref{sec:sensitivity} evaluates local sensitivity
around this nominal setting rather than using it for tuning.

\subsection{Closed-Loop Performance}
\label{sec:closed_loop}

\begin{figure*}[t]
    \centering
    \includegraphics[width=0.8\textwidth]{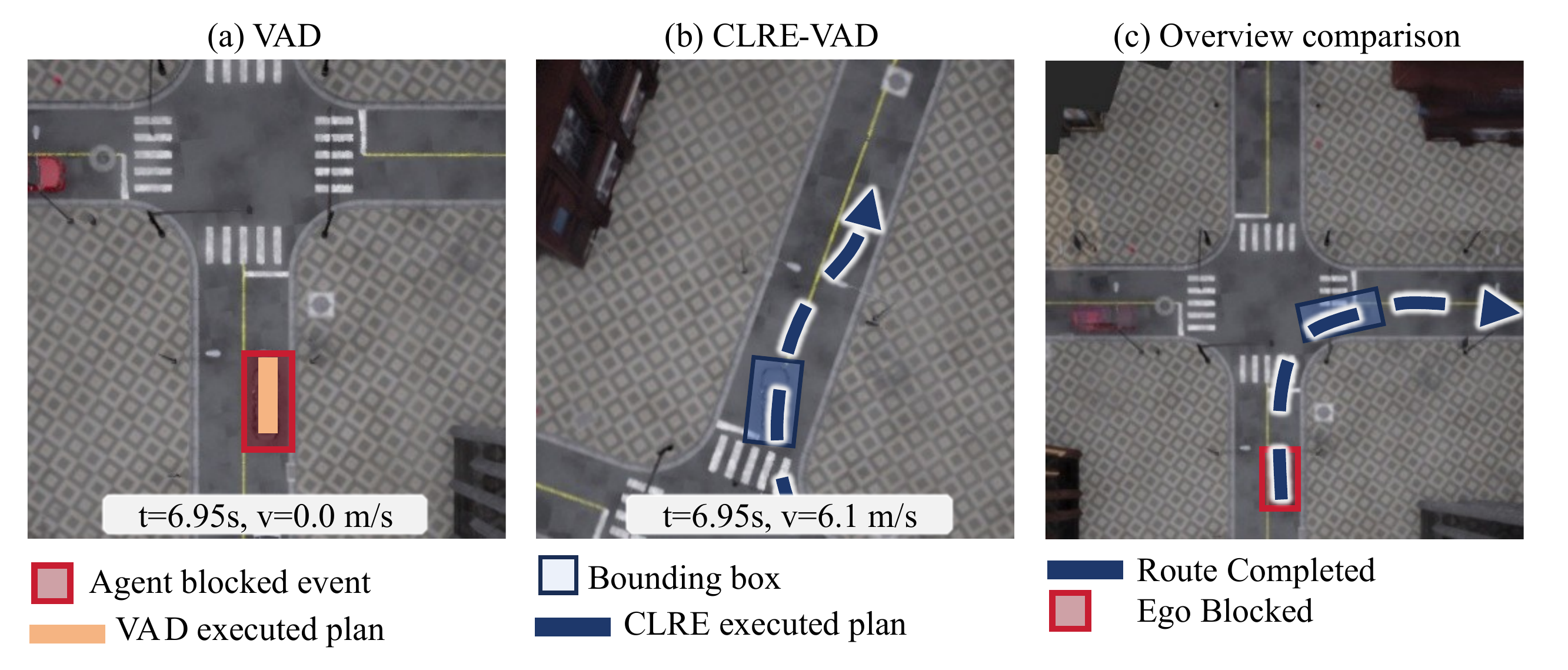}
    \caption{Effect of refinement on Route 16390. (a) Executing the nominal VAD
    trajectory directly, the vehicle remains at its spawn point with a
    zero-length plan and full brake applied at $t=6.95$~s. (b) At the same instant, CLRE is
    traveling at $6.1$~m/s. (c) Executed paths over the whole route: CLRE
    completes it, whereas the baseline covers no distance and stalls until timeout.}
    \label{fig:refinement}
\end{figure*}

\begin{figure}[!b]
    \centering
    \includegraphics[width=\columnwidth]{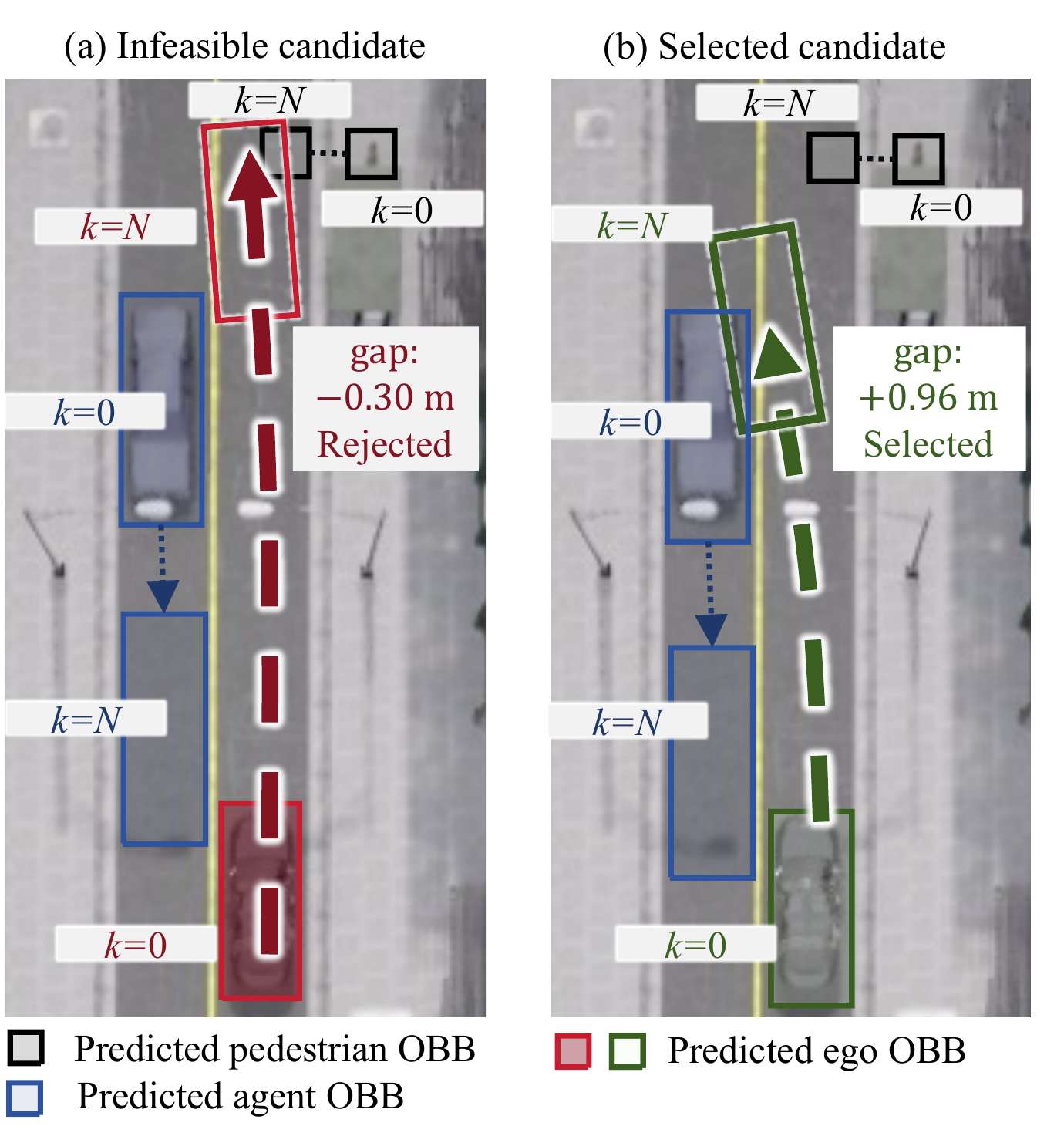}
    \caption{Prediction-conditioned OBB feasibility test on Route 17749, shown
    with the predicted agent, pedestrian, and ego boxes at $k=0$ and at
    $3$~s ($k=N$). (a) The lowest-cost candidate overlaps the predicted
    pedestrian box and is rejected, with
    $\operatorname{gap}=-0.30$~m. (b) The lowest-cost feasible candidate is
    selected, with $\operatorname{gap}=+0.96$~m.}
    \label{fig:OBB}
\end{figure}

Table~\ref{tab:overall_closed_loop} summarizes the closed-loop results.
On the 220-route VAD cohort, CLRE improves DS and RC by $13.63$ and
$15.83$ points, respectively, with paired bootstrap 95\% confidence
intervals of $[10.10,17.20]$ and $[12.01,19.86]$ and $p<10^{-12}$ by
sign tests. Collision events decrease from $117$ to $97$, or from $8.19$
to $5.62$ per kilometer driven. Figures~\ref{fig:refinement} and~\ref{fig:OBB} illustrate two representative
mechanisms.

\rev{Having established CLRE's effectiveness with VAD, we further evaluate its transferability to other upstream planners, UniAD~\cite{hu2023planning} and DriveTransformer~\cite{jia2025drivetransformer}, sharing the same CLRE parameters. CLRE improves DS and reduces collisions for both models, and improves RC for UniAD.
These results provide preliminary evidence of transfer across upstream
planners.}

\begin{table}[!t]
\vspace*{6pt}
\centering
\caption{Overall closed-loop performance.}
\label{tab:overall_closed_loop}
\footnotesize
\setlength{\tabcolsep}{3pt}
\begin{tabular}{lccccc}
\hline
Method
& Routes
& DS $\uparrow$
& RC $\uparrow$
& SR $\uparrow$
& Coll. $\downarrow$ \\
\hline

VAD~\cite{jiang2023vad}
& 220
& 42.26
& 55.69
& 20.91\%
& 117 \\

CLRE-VAD
& 220
& \textbf{55.89}
& \textbf{71.51}
& \textbf{28.64\%}
& \textbf{97} \\

\hline

UniAD~\cite{hu2023planning}
& 11
& 30.34
& 58.97
& 0.00\%
& 17 \\

CLRE-UniAD
& 11
& \textbf{40.81}
& \textbf{61.00}
& \textbf{9.09\%}
& \textbf{9} \\

\hline

DriveTransformer~\cite{jia2025drivetransformer}
& 11
& 39.20
& \textbf{85.40}
& 9.09\%
& 20 \\

CLRE-DriveTransformer
& 11
& \textbf{52.30}
& 71.14
& 9.09\%
& \textbf{5} \\

\hline
\end{tabular}
\end{table}

\subsection{Planner Ablation}
\label{sec:planner_ablation}

We ablate the progress term \eqref{eq:prog} and the OBB feasibility test in Table~\ref{tab:ablation}.

Removing the progress term substantially reduces forward motion. RC falls by
$17.6$ points, the number of completed routes decreases from $22$ to $15$,
the mean speed drops from $2.43$ to $1.64$~m/s.
The ablated variant also produces fewer collisions, including after
normalization by distance driven. This reduction should be interpreted
cautiously, since stalled vehicles may not reach the conflict regions of a
scenario. Overall, the progress term is important for maintaining route
advancement.

Removing the OBB feasibility test leaves distance driven within $0.2\%$ of
the full method, but increases collisions from $10$ to $17$ and the collision
rate per kilometer by $70\%$. This shows that the feasibility test provides
an additional check beyond the soft interaction term~\eqref{eq:interaction}.

\subsection{Execution-Layer Ablation}
\label{sec:exec_ablation}
We remove the execution layer of Section~\ref{sec:method:controller},
and use the original
tracking controller alone while keeping the refinement layer unchanged. The execution
parameters are fixed at $d_{\min}=2.0$ m, $a_{\mathrm{br}}=4.0$ m/s$^2$, and
$T_r=0.4$ s.

To prevent stalled or near-stationary behavior from artificially appearing
smooth, we evaluate execution smoothness only over moving ticks
($v>0.5$~m/s), using the mean absolute change in brake command and the
95th percentile of longitudinal jerk. The jerk is computed from the
speedometer signal using the same differentiation settings as the
Bench2Drive comfort pipeline.

As shown in Table~\ref{tab:ablation}, the execution layer reduces mean brake
variation from $0.101$ to $0.042$ and the jerk percentile from $38.3$ to
$28.7$. Collisions per kilometer also decrease from $4.98$ to $3.18$, while
keeping other main metrics (DS, RC, and the number of completed routes) stable.

\begin{table}[!tb]
\vspace*{6pt}
\centering
\caption{Ablations of CLRE-VAD on the first 36 routes.}
\label{tab:ablation}
\footnotesize
\setlength{\tabcolsep}{3.5pt}
\begin{tabular}{lccccccc}
\hline
Variant & DS & RC & Compl. & Coll. & Coll./km & Brake & Jerk \\
 & $\uparrow$ & $\uparrow$ & $\uparrow$ & $\downarrow$ & $\downarrow$
 & $\downarrow$ & $\downarrow$ \\
\hline
\multicolumn{8}{l}{\textit{Planner ablation}} \\
Full & \textbf{63.57} & 75.22 & \textbf{22} & 10 & 3.18
 & \textbf{0.042} & \textbf{28.7} \\
w/o progress & 53.13 & 57.59 & 15 & \textbf{2} & \textbf{0.79} & -- & -- \\
w/o feas.\ test & 58.30 & 74.39 & 20 & 17 & 5.40 & -- & -- \\
\hline
\multicolumn{8}{l}{\textit{Execution-layer ablation}} \\
w/o exec.\ layer & 61.59 & \textbf{75.58} & \textbf{22} & 16 & 4.98
 & 0.101 & 38.3 \\
\hline
\end{tabular}
\end{table}

\subsection{Objective and Sensitivity Analysis}
\label{sec:sensitivity}

The nominal objective weights are
$(w_{\mathrm{pt}},w_{\mathrm{ct}},w_{\mathrm{prog}})
=(0.05,1.2,1.0)$,
$(w_{\mathrm{out}},w_{\mathrm{in}},w_{\mathrm{bnd}},w_{\mathrm{cor}})
=(1.6,25,6,0.5)$, and
$(w_v,w_{\mathrm{sm}})=(0.6,0.8)$.
Because the objective terms have different numerical scales and aggregation
forms, the weight magnitudes are not directly comparable; they are chosen to
balance the resulting weighted contributions according to the intended
control priorities.

The progress term is the largest weighted contributor in $94.0\%$ of moving
solves, while the inner interaction term becomes more prominent when no
candidate is feasible.

\begin{figure}[!tb]
    \centering
    \includegraphics[width=1.0\columnwidth]{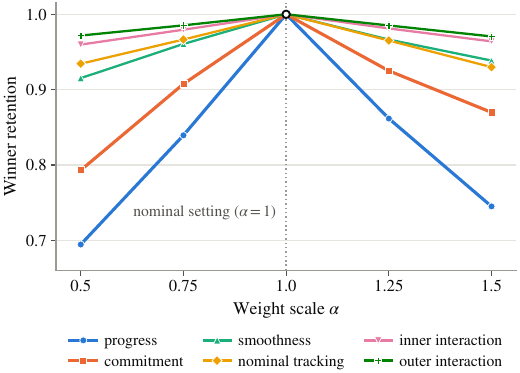}
    \caption{Winner retention under single-weight scaling over $41{,}858$
    development solves with at least two distinct feasible candidates.
    Retention is the fraction of solves whose selected
    candidate is unchanged from the nominal setting $\alpha=1$.}
    \label{fig:weight_sensitivity}
\end{figure}

We then scale six representative weights individually by
$\alpha\in\{0.5,0.75,1,1.25,1.5\}$ and recompute candidate selection.
As shown in Fig.~\ref{fig:weight_sensitivity}, progress and commitment are the
most sensitive terms, with winner retention of $0.69$ and $0.79$ when halved,
while the remaining weights retain the selected candidate in over $0.9$ of
solves across the tested range. No weight exhibits a narrow operating range;
systematic tuning beyond the current heuristic design remains future work.

\subsection{Numerical Solver Analysis}
\label{sec:solver_analysis}

Table~\ref{tab:solver_robustness} compares the deployed configuration of
Section~\ref{sec:experimental_setup} with $2\times$ and $5\times$ iteration
budgets. Candidate change is
the fraction of windows whose selected candidate differs from the deployed
solve; feasibility agreement is the fraction whose feasibility verdict is
unchanged. \rev{The two deviation columns are medians over windows in which both solves return a feasible candidate.}

\begin{table}[!tb]
\centering
\caption{Solver sensitivity over 3275 replanning windows.}
\label{tab:solver_robustness}
\footnotesize
\setlength{\tabcolsep}{3pt}
\begin{tabular}{lcccc}
\hline
& Candidate & Feasibility & \rev{Dev.\ at} & \rev{Max dev.} \\
Comparison & change & agreement & \rev{$k=2$ (m)} & \rev{(m)} \\
\hline
$2\times$ iteration budget & 51.8\% & 99.97\% & \rev{0.005} & \rev{0.11} \\
$5\times$ iteration budget & 58.7\% & 99.91\% & \rev{0.006} & \rev{0.13} \\
\hline
\end{tabular}
\end{table}

\rev{The results show that larger iteration budgets frequently change the selected candidate, while
feasibility remains nearly unchanged. However, the discrepancy is concentrated near
the horizon end: among windows in which the selected candidate changes,
the largest deviation occurs at $k=N$ in $74\%$ and $76\%$ of
cases for the $2\times$ and $5\times$ budgets, respectively. The first
executed waypoint differs by only $14$~mm in median under both budgets,
while the corresponding commanded-speed differences are $0.07$ and
$0.08$~m/s. The multi-restart
reference search agrees with the deployed feasibility verdict in $99.54\%$ of
windows, and raising the clearance threshold from $0.05$ to $0.30$~m changes
the selected candidate in only $7.2\%$ of them.}

\rev{The current implementation is not real-time at 20~Hz. On an RTX~4090 GPU and an i9-13900KF CPU, the forward pass of VAD takes
$132.1$~ms per control step on average over the 220-route cohort, and CLRE
planning adds $100.2$~ms on the CPU.} Since Bench2Drive is evaluated in synchronous CARLA,
computation time affects wall-clock simulation speed but not the simulated
control timing. \rev{
On the development windows, vectorizing the feasibility test reduces its
offline runtime from $28.9$~ms to $0.64$~ms with identical decisions. GPU-parallel trajectory optimization has also been demonstrated for
autonomous-vehicle MPC~\cite{lee2024gpu}. Since the candidate refinements in
CLRE are independent, a GPU-native batched implementation could exploit this
parallelism and could further scale to higher-throughput or multiple GPUs.
We leave such
hardware-parallel real-time implementation to future work.}
\section{Conclusion}
\label{sec:conclusion}
In this paper, we presented CLRE, a training-free framework for improving the
closed-loop behavior of frozen learned trajectory planners. CLRE combines a
receding-horizon refinement method that balances nominal-path tracking, route
progress, and predicted-agent interaction with a prediction-conditioned OBB
feasibility test that screens refined candidates against predicted agent
motion. It further introduces an execution layer with a forward-range speed
bound, route-centerline backup, and saturated braking for closed-loop
execution. Bench2Drive experiments demonstrated improved closed-loop
performance with VAD and provided preliminary evidence of transfer to other
upstream planners without retuning.

The present study is limited to simulation, and the feasibility test depends on
predicted agent motion. Future work will address multimodal and
uncertainty-aware feasibility screening, broader cross-backbone evaluation,
and GPU-parallel real-time implementation.
\section*{Disclosure of Large Language Model Use}

OpenAI ChatGPT~\cite{OpenAIChatGPT} and Claude Code~\cite{AnthropicClaudeCode}
were used to support literature search, coding, and language editing. The authors reviewed and validated all material incorporated into the manuscript and take full responsibility for the content of the paper.
\FloatBarrier
\bibliographystyle{IEEEtran}
\bibliography{references}

\end{document}